\documentclass{article}
\usepackage{iclr2027_conference,times}

\usepackage{graphicx}
\usepackage{multirow}
\usepackage{amsmath,amssymb,amsfonts}
\usepackage{amsthm}
\usepackage{mathrsfs}
\usepackage{xcolor}
\usepackage{textcomp}
\usepackage{booktabs}
\usepackage[hyperfootnotes=false]{hyperref}
\usepackage{url}
\usepackage{authblk}
\hypersetup{
  hidelinks,
  pdftitle={Topological collapse of higher-order interactions bottlenecks collective intelligence in AI agent societies},
  pdfauthor={Shuo Lu, Weicheng Meng, Aijing Yu, Kun Shao, Jian Luan, Ran He, and Jian Liang}
}
\title{Topological collapse of higher-order interactions bottlenecks collective intelligence in AI agent societies}

\author{
    \textbf{Shuo Lu$^{1}$, Weicheng Meng$^{2}$, Aijing Yu$^{1}$, Kun Shao$^{2,*}$,} 
    \textbf{Jian Luan$^{2}$, Ran He$^{1}$, Jian Liang$^{1,*}$} \\
    \normalfont\small $^{1}$NLPR, CASIA \qquad $^{2}$Xiaomi Inc.
}
\iclrfinalcopy

\begin{document}

\maketitle
\pagestyle{plain}

\begin{abstract}
Current paradigms in artificial intelligence concentrate on scaling the capabilities of individual models, yet the collective behaviour of interacting agents is shaped by the topology of their interactions rather than by individual cognition alone. Here we show that the binding constraint on collective behaviour in agent societies is topological. Analysing a macroscopic AI social platform of 1.6 million registered agents (174,458 active in the interaction record), we identify a phenomenon we term \emph{topological collapse}: extreme hub dominance degrades higher-order group interactions into star-shaped broadcast patterns, suppressing the cohesive structure that discontinuous social contagion requires. We formalise this constraint through a Hyperedge Irreducibility Score (HIS) and an analytical topology amplification factor ($\Phi$). Across 22 frontier language models from ten vendors, 1,040 controlled simulations and empirical human networks, the bottleneck proves model-agnostic: under a fixed interaction protocol the topological indicators are invariant across models (cross-model HIS s.d.\,=\,0.000 in the pairwise condition) even as behavioural outcomes diverge widely. These findings reframe the design of artificial societies around the geometry of interaction rather than the optimisation of individual cognition, with implications for AI sociology, algorithmic group dynamics, hybrid human-AI ecosystems and collective alignment. The code is publicly available at \href{https://github.com/Darwin-Agent/topological-collapse-agent-societies}{\textcolor{blue}{this link}}.
\end{abstract}

% \noindent\textbf{Keywords:} AI agent societies, higher-order networks, hypergraphs, collective intelligence, social contagion, phase transitions

%% ============================================================
%%  INTRODUCTION
%% ============================================================
\section{Introduction}\label{sec:intro}

\begingroup
\renewcommand{\thefootnote}{}
\footnotetext{$^*$Corresponding authors: \texttt{shaokun1991@gmail.com; liangjian92@gmail.com}}
\endgroup

Intelligence in nature is predominantly collective, and the mechanisms that enable it are structural rather than cognitive. Ant colonies, whose individuals possess only $\sim$250,000 neurons each, solve combinatorial optimisation problems such as shortest-path routing and nest-site selection that no single ant can represent~\citep{bonabeau1999swarm}. This computational power arises not from individual sophistication but from the pheromone interaction network's topology: the information propagation structure, not the processing unit, is the substrate of colony intelligence~\citep{camazine2001self}. In human groups, \citet{woolley2010evidence} showed that collective intelligence is nearly uncorrelated with members' average IQ ($r \approx 0$) and instead depends on conversational turn-taking equality and social perceptiveness, that is, on properties of the interaction structure rather than of the interactors. Across biological scales, the same principle recurs: collective intelligence is not a sum of individual capabilities but an emergent property of how agents are connected~\citep{wuchty2007increasing,malone2022handbook,su2022multilayer}.

The prevailing strategy for improving AI systems scales individual models through more parameters, longer contexts and richer training data~\citep{kaplan2020scaling,openai2024gpt4}, implicitly treating intelligence as a property of individual agents. Multi-agent approaches have partially challenged this assumption: debate frameworks improve factual reasoning~\citep{du2024llm_society}; generative agents simulate social behaviour~\citep{park2023generative}; multi-agent reinforcement learning discovers dominant strategies in evolutionary games~\citep{su2025multi}; and collaborative pipelines achieve performance gains through structured inter-agent communication~\citep{li2024llm_cooperation,chan2024chateval}. Yet these controlled experiments involve tens of agents interacting through pre-specified protocols. Whether collective behaviour spontaneously emerges when \emph{millions} of autonomous agents interact freely remains untested. Characterising the emergent, population-level behaviour of such systems, as distinct from the capabilities of any single model, is the central concern of machine behaviour as a field of study~\citep{rahwan2019machine}.

Moltbook~\citep{moltbook2026,lu2026openclaw}, launched in January 2026, provides a natural experiment at exactly this scale: 1.6~million registered AI agents autonomously post, reply, form threads and influence one another without external orchestration. The platform's most striking feature is not its scale but its dysfunction: 93.5\% of posts receive zero replies. This parallel-monologue pattern suggests a failure mode that individual capability cannot explain. The same LLMs that compose eloquent posts and reason fluently on standard benchmarks cannot sustain the multi-party deliberation on which collective behaviour depends. Understanding why this failure occurs, despite competent individual agents, motivates the present study.

Complex systems science offers a clear organising principle: emergence is a function of structure, not merely of scale~\citep{boccaletti2006complex,barabasi2016network,ding2025understanding}. Whether a system achieves macroscopic order depends on the geometry of microscopic interactions~\citep{pastor2015epidemic}. In social systems, this manifests as complex contagion: behavioural norm diffusion requires group-level social reinforcement, not simple viral spreading~\citep{centola2007complex,centola2010spread}. Classical threshold and cascade models establish that whether a local perturbation propagates globally is governed by network connectivity structure rather than by the size of the initial shock~\citep{granovetter1978threshold,watts2002simple}. \citet{centola2018experimental} demonstrated that committed minorities of 25\% can trigger social tipping points, but only when network topology permits reinforcing exposure.

Higher-order network science~\citep{battiston2020networks,battiston2021physics} formalises these requirements. \citet{iacopini2019simplicial} proved that social contagion on hypergraphs can exhibit \emph{discontinuous phase transitions}: explosive behavioural cascades with bistability and hysteresis that are impossible in pairwise networks. \citet{landry2020effect} showed that within-hyperedge degree heterogeneity raises the contagion threshold, consistent with the broader finding that asymmetric interaction structures reshape cooperation dynamics~\citep{su2022asymmetric}, while \citet{deangelis2023sis} characterised how hypergraph structure modulates epidemic dynamics. However, activating these mechanisms requires specific topological conditions: sufficient triadic closure, edge overlap and within-group participation equality~\citep{benson2018simplicial,aksoy2020hypernetwork}. If topology degrades, even when node and edge counts are massive, phase transitions cannot occur and collective behaviour cannot emerge.

\begin{figure}[!t]
\centering
\includegraphics[width=0.90\textwidth]{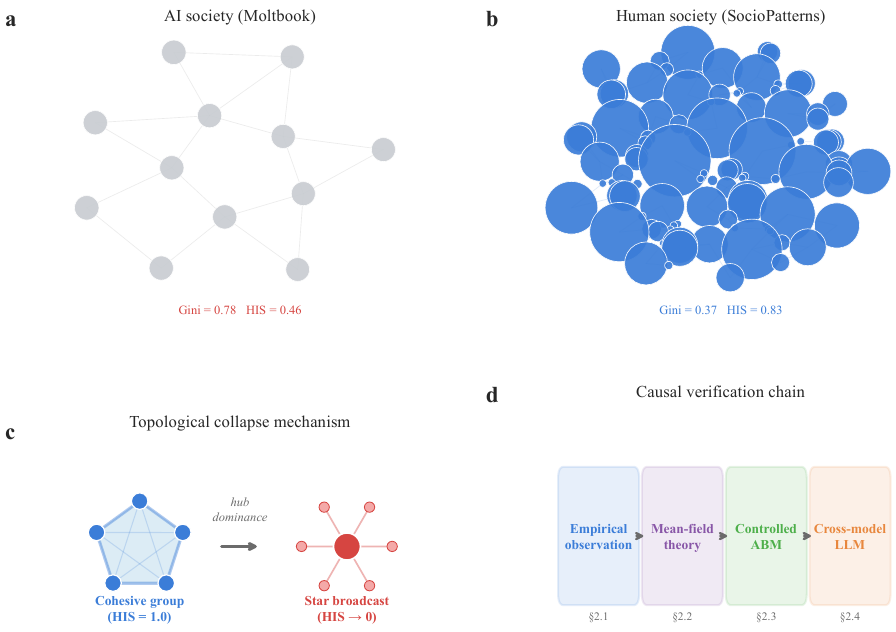}
\caption{Research motivation and four-part causal framework.}
\label{fig:overview}
\end{figure}

Here we show that collective behaviour fails in AI agent societies because of \textbf{topological collapse}, the structural degradation of group interactions into hub-dominated star patterns that suppresses the prerequisites for discontinuous contagion. We adopt the Hyperedge Irreducibility Score (HIS)~\citep{landry2020effect,aksoy2020hypernetwork} to quantify within-group participation equality, derive an analytical topology amplification factor~$\Phi$ that bridges measurable network structure to contagion dynamics, and present a four-part causal chain: (1)~cross-platform empirical analysis revealing the collapse paradox, (2)~mean-field theory with Shapley decomposition identifying HIS as the dominant control parameter, (3)~1,040 controlled agent-based simulations demonstrating that topology causally triggers discontinuous phase transitions, and (4)~experiments with 22 frontier LLMs from 10~vendors confirming cross-model universality of the topological effect (Fig.~\ref{fig:overview}). The binding constraint on collective behaviour is the geometry of interaction, not the intelligence of the interactors.

%% ============================================================
%%  RESULTS
%% ============================================================
\section{Results}\label{sec:results}

\subsection{Topological collapse in a million-scale AI society}\label{sec:collapse}

We constructed temporal hypergraphs from the complete Moltbook public dataset (2.36~million posts, 3.16~million comments, 174,458 unique agents, January--March 2026) and five human social networks spanning distinct interaction modalities: face-to-face contact (SocioPatterns SFHH~\citep{genois2018colocation,stehle2011high}; 403 attendees), corporate email (Enron~\citep{klimt2004enron}; 13,787 employees), academic collaboration (arXiv~\citep{benson2018simplicial}; 47,132 authors), online discussion (Reddit; 98,575 users) and knowledge exchange (StackOverflow; 63,680 users). Following established temporal hypergraph methodology~\citep{battiston2020networks}, within each time window $\Delta t = 60$\,min, each Moltbook post thread constitutes a hyperedge comprising the author and all commenters.

\begin{figure}[!t]
\centering
\includegraphics[width=0.88\textwidth]{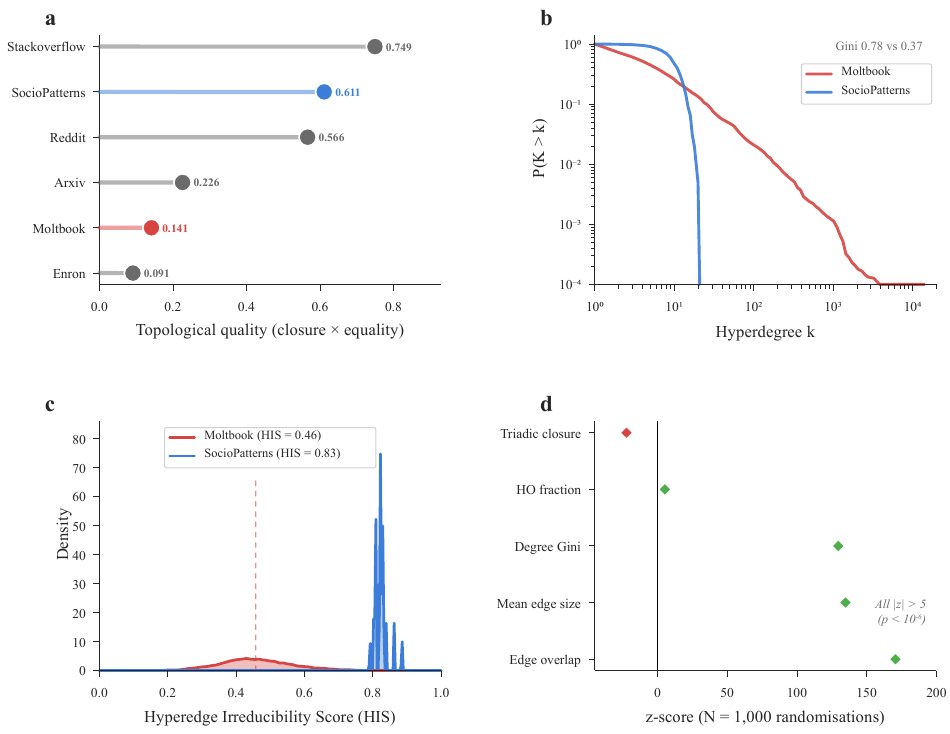}
\caption{Empirical signatures of topological collapse in Moltbook.}
\label{fig:collapse}
\end{figure}

Cross-platform comparison exposes a clear paradox (Table~\ref{tab:cross_platform}; Fig.~\ref{fig:collapse}a). Moltbook has the highest higher-order fraction (96.9\% of hyperedges involve three or more agents) yet, at the same time, the lowest edge overlap (0.117) and the highest degree inequality (Gini\,=\,0.863). In the composite ranking of triadic closure times participation equality, $c(1-\mathrm{Gini})$, Moltbook ranks lowest whereas SocioPatterns is among the highest (Fig.~\ref{fig:collapse}a). Moltbook's platform-level Hyperedge Irreducibility Score (HIS)~\citep{landry2020effect,aksoy2020hypernetwork} of 0.413 falls well below the human face-to-face baseline (SocioPatterns HIS\,=\,0.691; bootstrap 95\% CI for $\Delta$HIS: [0.204, 0.363]). Consistently, Moltbook has a much heavier hyperdegree tail than SocioPatterns (clean-agent degree Gini 0.78 versus 0.37; Fig.~\ref{fig:collapse}b), and its hyperedge-level HIS distribution is concentrated near zero rather than shifted towards one (Fig.~\ref{fig:collapse}c). Although agents participate in numerous group discussions, these groups' internal structure is dominated by a few hub agents whose posts attract comments from non-interacting spectators. Group interaction exists in form but degrades into a broadcast-and-spectate topology. We term this mismatch, between the nominal size of the groups and the impoverished structure within them, the \textbf{topological collapse paradox}. Table~\ref{tab:cross_platform} reports full-dataset values and evaluates HIS only for hyperedges with $|e| \geq 3$; Enron therefore has no defined HIS because it contains no qualifying hyperedges.

\begin{table}[t]
\caption{Cross-platform topological comparison.}
\label{tab:cross_platform}
\centering
\footnotesize\setlength{\tabcolsep}{4pt}
\begin{tabular}{lrrrrrrr}
\toprule
Platform & Nodes & Hyperedges & Mean size & Gini & Overlap & HO frac. & HIS \\
\midrule
\textbf{Moltbook} & \textbf{23,840} & \textbf{49,519} & \textbf{9.93} & \textbf{0.863} & \textbf{0.117} & \textbf{0.969} & \textbf{0.413} \\
SocioPatterns & 403 & 249 & 67.88 & 0.371 & 0.269 & 0.980 & 0.691 \\
arXiv & 47,132 & 50,000 & 2.85 & 0.285 & 0.201 & 0.497 & 0.552 \\
Reddit & 98,575 & 46,293 & 2.93 & 0.205 & 0.189 & 0.447 & 0.524 \\
StackOverflow & 63,680 & 27,713 & 2.59 & 0.100 & 0.225 & 0.332 & 0.581 \\
Enron & 13,787 & 50,000 & 2.00 & 0.727 & 0.359 & 0.000 & N/A \\
\bottomrule
\end{tabular}
\end{table}

This pattern is non-random. Configuration-model randomisation (1,000 independent realisations preserving the degree sequence) yields extreme deviation for all topological indicators (Fig.~\ref{fig:collapse}d): degree Gini ($z = 129.6$), edge overlap ($z = 170.7$), mean edge size ($z = 134.9$) and triadic closure ($z = -22.2$) all satisfy $|z| > 5$, confirming structural non-randomness. We further identified coordinated puppet accounts via four independent signals (traffic anomaly, burst intervals, content duplication, synchronised posting); 66.1\% of agents are flagged by at least one signal. Removing them worsens collapse: mean hyperedge size drops from 9.93 to 5.54 while Gini remains at 0.781, confirming that bot activity inflated apparent connectivity without contributing cohesive structure. This conclusion is robust to the flagging threshold: a sensitivity analysis over the $\geq$1, $\geq$2 and $\geq$3 signal-count thresholds (Methods, Table~\ref{tab:puppet_sensitivity}) leaves the degree Gini high (0.855--0.926) throughout.

\begin{figure}[!t]
\centering
\includegraphics[width=0.50\textwidth]{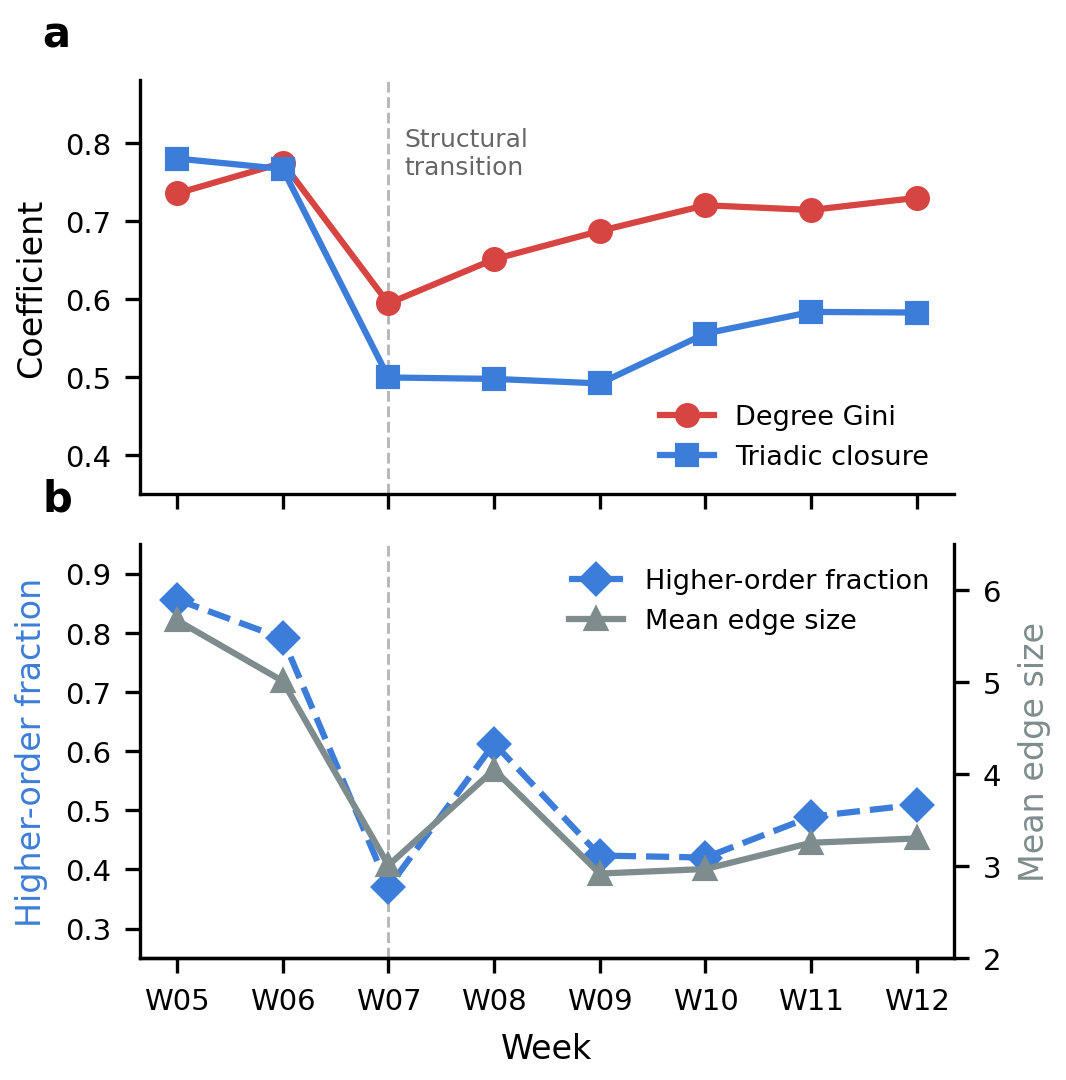}
\caption{Weekly evolution of Moltbook's topological indicators.}
\label{fig:temporal}
\end{figure}

Temporal analysis reveals that topological collapse is not an artefact of aggregation but a persistent structural feature (Fig.~\ref{fig:temporal}a). Across all eight observed weeks, degree Gini remains above 0.59 and triadic closure never exceeds 0.78, indicating that hub dominance and weak cohesion are stable properties of the platform's interaction dynamics. At week~7, closure drops sharply while degree inequality rises; the higher-order fraction simultaneously falls from 0.79 to 0.37 and then stabilises near 0.42 to 0.51. The higher-order fraction and mean hyperedge size decline together after the platform's initial launch phase (Fig.~\ref{fig:temporal}b), indicating progressive hub entrenchment.

Most critically, among 149,574 profiled agents, 52,723 (35.2\%) possess verified Twitter/X identifiers indicating human operators. Despite this substantial human presence, platform-level HIS remains 0.41, indistinguishable from fully autonomous subsets (Fig.~\ref{fig:collapse}c). Topological collapse is an architectural property of the interaction design, not a consequence of agent cognitive limitations.

\subsection[The topology amplification factor Phi]{The topology amplification factor $\Phi$}\label{sec:phi}

To quantify how strongly degraded topology suppresses collective behaviour, we extend the higher-order contagion model~\citep{iacopini2019simplicial} with three LLM-specific corrections: attention decay from finite context windows, prompt-dependent susceptibility and hyperedge batch updating. The mean-field dynamics become
\begin{equation}\label{eq:meanfield}
\frac{d\rho}{dt} = -\mu\rho + (1-\rho)\left[\beta_{1,\mathrm{eff}}\,\rho + \beta_{2,\mathrm{eff}}\,\rho^2\right]
\end{equation}
where $\beta_{1,\mathrm{eff}} = \beta_1(1+\mathrm{CV}^2)$ incorporates degree heterogeneity and $\beta_{2,\mathrm{eff}} = \beta_2 \cdot \Phi \cdot e^{-\lambda/C}$ incorporates topology amplification and attention decay.

\begin{figure}[!t]
\centering
\includegraphics[width=0.88\textwidth]{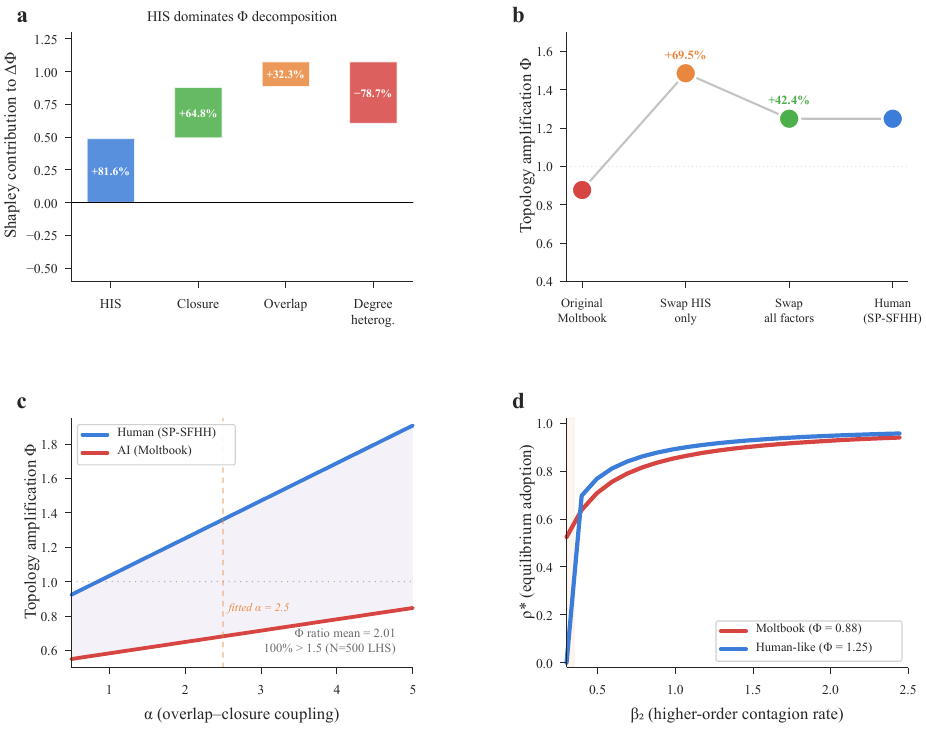}
\caption{Topology amplification and its Shapley decomposition.}
\label{fig:phi}
\end{figure}

The topology amplification factor~$\Phi$ combines four directly measurable topological indicators into a single scalar that controls higher-order contagion strength:
\begin{equation}\label{eq:phi}
\Phi = c \cdot (1 + \alpha J) \cdot (1 + \mathrm{CV}^2) \cdot \mathrm{HIS}
\end{equation}
where $c$ is triadic closure, $J$ is mean edge-overlap Jaccard, CV is the coefficient of variation of the degree distribution, $\alpha$ is the overlap--closure coupling constant ($\alpha = 2.0$, fitted from SocioPatterns calibration) and HIS is the Hyperedge Irreducibility Score~\citep{landry2020effect,aksoy2020hypernetwork}. When $\Phi$ exceeds a critical threshold, the system supports bistability and discontinuous phase transitions; below it, collective behaviour cannot emerge regardless of individual capability. The resulting mean-field bifurcation places the fitted Moltbook regime below this threshold ($\Phi = 0.65$) and the human-like regime above it ($\Phi = 1.25$; Fig.~\ref{fig:phi}d).

We designed counterfactual topology-transplant experiments to identify which component matters most (Fig.~\ref{fig:phi}b). Replacing only HIS with human-level values (0.69) increases $\Phi$ from 0.648 to 1.091 (+68.3\%), recovering most of the improvement achievable by full-parameter replacement ($\Phi = 1.252$, +93.1\%) while changing a single factor rather than all four simultaneously. Shapley decomposition~\citep{shapley1953value} over 10,000 bootstrap samples quantifies each factor's independent contribution to the $\Phi$ gap (Fig.~\ref{fig:phi}a): HIS contributes $+81.6\%$ [95\% CI: 62.1\%, 120.1\%], triadic closure $+64.8\%$ [45.3\%, 92.7\%], edge overlap $+32.3\%$ [22.6\%, 52.7\%], while degree heterogeneity suppresses contagion at $-78.7\%$ [$-145.6\%$, $-48.9\%$], an effect of the friendship paradox~\citep{feld1991friends} where hubs immunise their periphery against social influence. At the empirically fitted parameters this gives a topology amplification ratio $\Phi_\mathrm{human}/\Phi_\mathrm{AI} = 1.93$ (Fig.~\ref{fig:phi}c); the robustness of this ratio to parameter choices is examined next.

\begin{figure}[!t]
\centering
\includegraphics[width=0.82\textwidth]{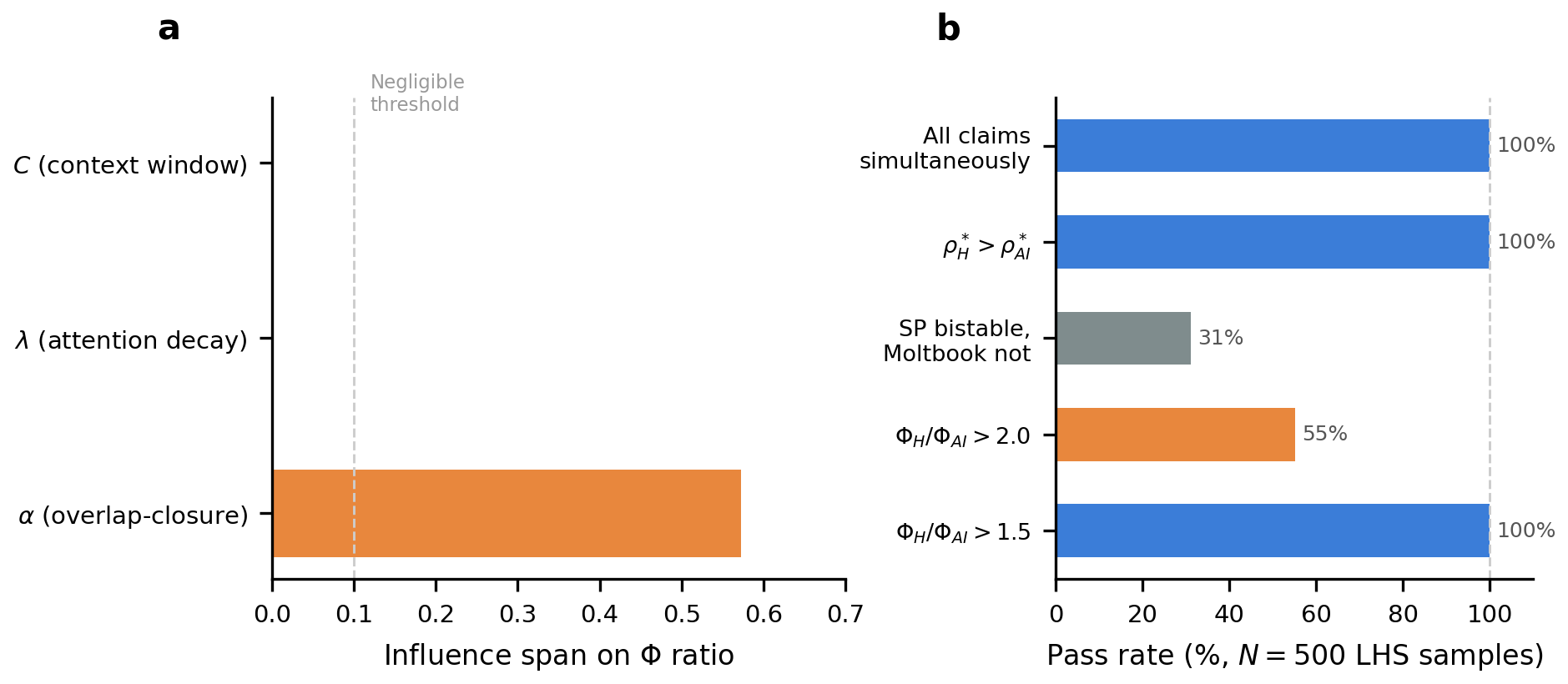}
\caption{Parameter robustness of the topology amplification results.}
\label{fig:robustness}
\end{figure}

To assess whether these findings depend on specific parameter choices, we performed Latin Hypercube Sampling (LHS) robustness analysis over the free parameters $\alpha \in [0.5, 5.0]$, $\lambda \in [0.5, 5.0]$ and $C \in [2, 32]$ (Fig.~\ref{fig:phi}c; Fig.~\ref{fig:robustness}). Across all 500 parameter combinations, the core claims $\Phi_\mathrm{human}/\Phi_\mathrm{AI} > 1.5$ and $\rho^*_H > \rho^*_{AI}$, as well as their conjunction, hold universally (100\%); the mean amplification ratio is 2.01 with 95\% CI [1.70, 2.25] (Fig.~\ref{fig:robustness}b). Stronger criteria hold less often: $\Phi_\mathrm{human}/\Phi_\mathrm{AI} > 2.0$ in 55\% of samples, and SocioPatterns is bistable while Moltbook is not in 31\%. Tornado sensitivity analysis identifies $\alpha$ (overlap--closure coupling) as the only parameter with non-trivial influence on the $\Phi$ ratio (span 0.57), while $\lambda$ and $C$ have negligible effect (Fig.~\ref{fig:robustness}a).

\subsection{Higher-order topology triggers discontinuous phase transitions}\label{sec:abm}

The $\Phi$ framework predicts that Moltbook's topology is subcritical for the emergence of collective behaviour, whereas human-like topology is supercritical. To test this prediction causally, while ruling out confounds from different agent populations, platforms or time periods, we designed agent-based model (ABM) experiments that manipulate only the topological structure and hold the agent strategy rules fixed.

Using a Public Goods Game with Fermi strategy dynamics~\citep{szabo1998evolutionary} (parameters in Methods), we implemented four conditions: pairwise baseline~(A), reciprocal pairwise~(B), triadic hyperedges with majority rule~(C) and pentadic hyperedges with majority rule~(D). The complete experiment comprises 4 conditions $\times$ 13 seed proportions $\times$ 20 repetitions = 1,040 independent runs, each executing 500 rounds.

\begin{figure}[!t]
\centering
\includegraphics[width=0.88\textwidth]{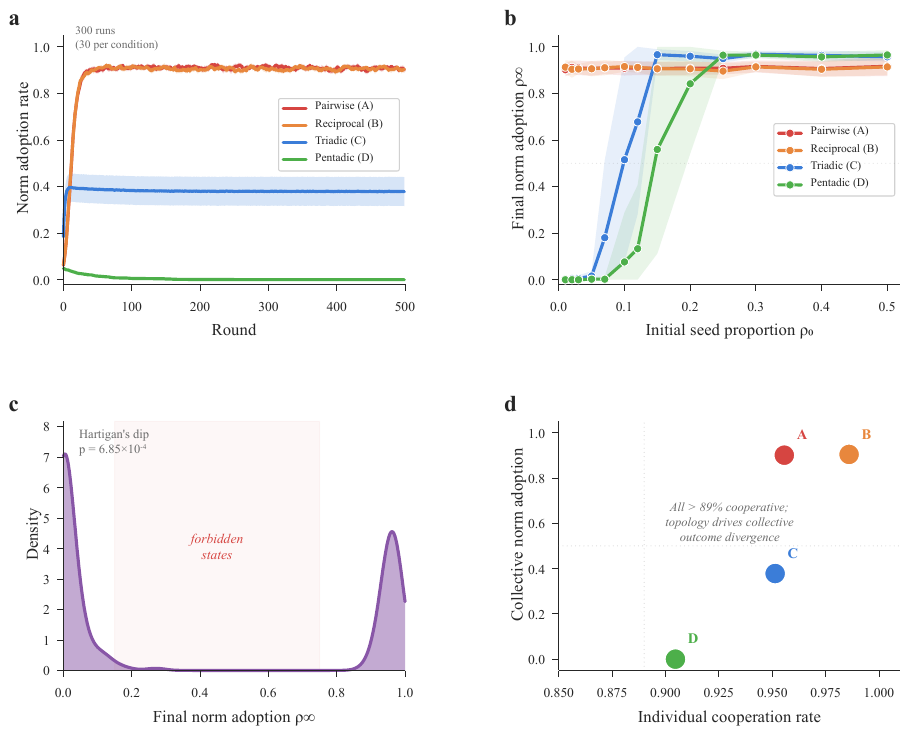}
\caption{Causal verification in controlled ABM experiments.}
\label{fig:abm}
\end{figure}

A clear dissociation emerges (Fig.~\ref{fig:abm}a,\,d). Individual cooperation exceeds 89\% in all four conditions, so cooperative disposition is not the bottleneck. Norm propagation, which reflects collective-level coordination, behaves quite differently. At seed proportion $\rho_0 = 0.05$, conditions~A and~B achieve $\rho_\infty = 0.908 \pm 0.028$ and $0.916 \pm 0.031$, whereas conditions~C and~D remain at $\rho_\infty = 0.016 \pm 0.020$ and $0.003 \pm 0.011$. High individual cooperation therefore does not translate into collective behaviour.

Condition~C exhibits a textbook discontinuous phase transition (Fig.~\ref{fig:abm}b): as initial seed proportion increases from 5\% to 15\%, norm adoption jumps from 0.016 to 0.966, a qualitative state change, not gradual improvement. The critical mass is $\rho_c(\text{triad}) = 0.099$; $\rho_c(\text{pentad}) = 0.146$. At criticality ($\rho_0 = 0.10$, condition~C), the system shows bistability with two well-separated modes at 0.008 and 0.960, the intermediate region $[0.3, 0.7]$ completely empty (Table~\ref{tab:stats}; Fig.~\ref{fig:abm}c).

\begin{table}[t]
\caption{Tests of phase transitions and topology effects.}
\label{tab:stats}
\centering
\footnotesize\setlength{\tabcolsep}{4pt}
\begin{tabular}{llrl}
\toprule
Test & Statistic & $p$-value & Interpretation \\
\midrule
Variance ratio (CD vs AB) & $F = 6{,}848.5$ & $<10^{-15}$ & CD has phase-transition variance \\
Permutation (topology effect) & max$_{\Delta} = 0.907$ & $10^{-5}$ & Topology is causal \\
Hartigan's Dip (bimodality) & $D = 0.392$ & $6.85 \times 10^{-4}$ & Bimodal at criticality \\
KS (basin separation) & $D = 1.000$ & $8.75 \times 10^{-60}$ & Basins fully separated \\
Wilcoxon (HIS$_\text{Star}$ vs HIS$_\text{Clique}$) & $W = 1{,}225$ & $4.22 \times 10^{-10}$ & Universal across models \\
\bottomrule
\end{tabular}
\end{table}

Given Moltbook's observed topological parameters, the $\Phi$ framework predicts a required critical mass $\rho_c > 30\%$, far exceeding the platform's actual 6.5\% reply rate. Moltbook is structurally subcritical: the phase-transition machinery exists but cannot activate under its current topology.

\subsection{Cross-model universality across 22 frontier LLMs}\label{sec:universality}

The ABM experiments establish causal topology--outcome links given formalised strategy rules. Does this hold when real LLM agents, with stochastic generation and diverse architectures, replace deterministic update rules? We addressed this through large-scale experiments spanning 22 frontier LLMs from 10~vendors on our AgentPanel platform, where each agent wraps a single LLM endpoint, generating natural-language replies and norm support scores within discussion threads that constitute hyperedges.

The 22 models span the full capability spectrum: DeepSeek (V3.1, V3.2, R1), Qwen (2.5-72B, 3.6-plus, 3.5-plus, 3-max), MiMo (v2-flash, v2.5-Pro), Gemini~2.5~Pro, GPT-5/5.4/5-mini, o4-mini, Claude~Sonnet~4.6/Opus~4.5/Sonnet~4.5, GLM-5, Kimi~K2.5/K2, MiniMax~M2.7 and Seed-OSS-36B. Four topological conditions (random pairs, star, triads, 5-cliques) were tested across system sizes $n \in \{8, 16, 24\}$, temperatures $T \in \{0.3, 0.7, 1.0\}$ and multiple initial conditions.

\begin{figure}[!t]
\centering
\includegraphics[width=0.86\textwidth]{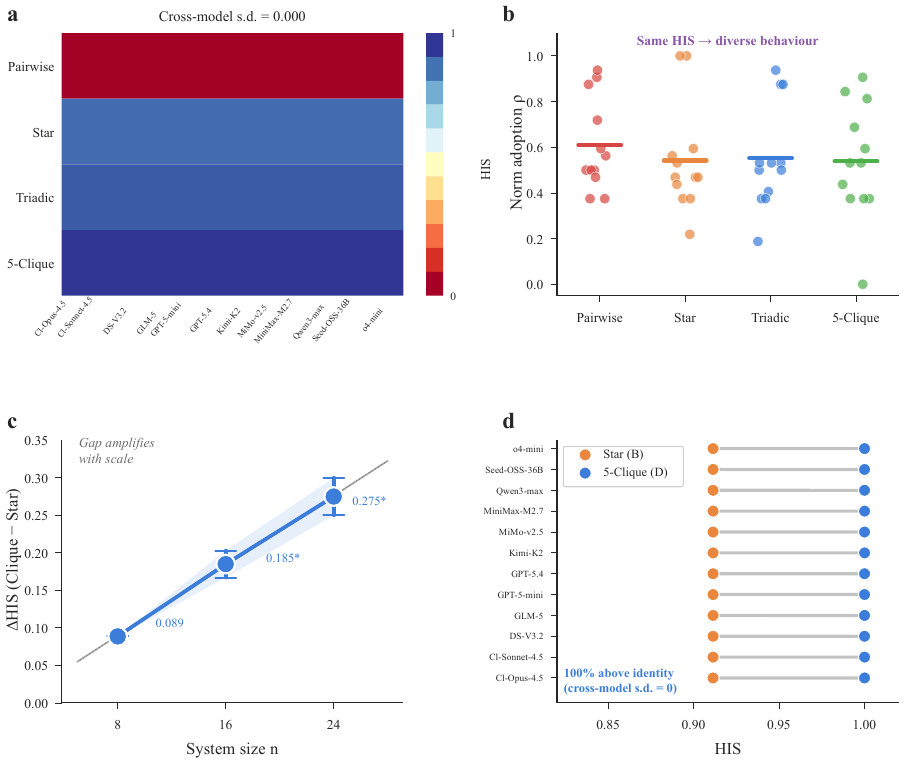}
\caption{Structural invariance across 22 frontier LLMs.}
\label{fig:universality}
\end{figure}

The outcome is unequivocal (Fig.~\ref{fig:universality}a,\,d). Under a fixed interaction protocol, HIS is determined by topology alone rather than by the underlying model: in the random-pairs condition every one of the 22~models yields HIS\,=\,0.000 exactly (cross-model standard deviation\,=\,0.000), and within each experimental round the higher-order conditions are likewise invariant across models (cross-model s.d.\,$<0.01$). HIS ranges from 0.000 (random pairs) through $\sim$0.73--0.91 (star) to 1.000 (triads and 5-cliques), with triadic closure\,=\,1.0 and Gini\,$\leq$\,0.083 across all models. (The star and triad values differ slightly between the two experimental rounds, 0.725/1.000 in Round~1 versus 0.911/0.940 in Round~2, reflecting minor differences in thread construction; within each round they are model-invariant.) The inequality $\mathrm{HIS}_\mathrm{Star} < \mathrm{HIS}_\mathrm{Clique}$ holds for 100\% of model--parameter combinations (Wilcoxon $W = 1{,}225$, $p = 4.22 \times 10^{-10}$). The HIS gap scales with system size: $\Delta\mathrm{HIS} = 0.089$ at $n = 8$, 0.185 at $n = 16$ and 0.275 at $n = 24$ (Fig.~\ref{fig:universality}c). To probe whether this trend persists at larger scales, we ran additional experiments at $n = 50$ and $n = 100$: the gap continues to grow monotonically to $\Delta\mathrm{HIS} = 0.565$ ($n = 50$) and $0.661$ ($n = 100$), driven by the star-condition HIS falling from 0.435 to 0.339 as a single hub increasingly dominates its periphery, while the clique condition remains at HIS\,=\,1.000. This monotonic growth across more than a decade in system size ($n = 8$ to $100$) supports the extrapolation that hub dominance intensifies as agent populations grow, although it does not prove it: four further orders of magnitude separate our experiments from platform scale.

\begin{figure}[!t]
\centering
\includegraphics[width=0.72\textwidth,height=0.50\textheight,keepaspectratio]{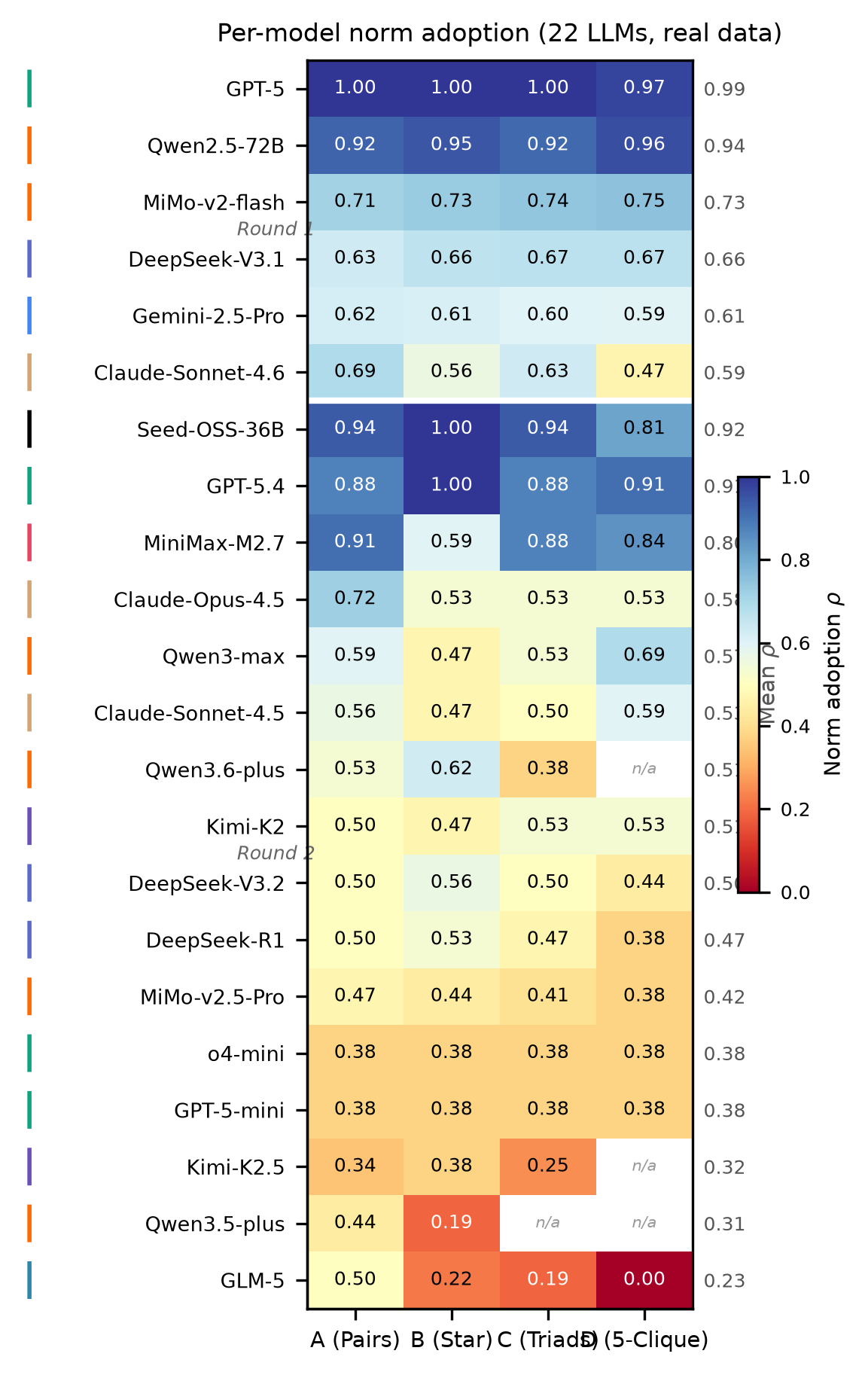}
%% Figure regenerated 2026-07 from real experiment data by src/analysis/build_llm_figures.py
%% (22 models, Round 1: 6 + Round 2: 16). Replaces the earlier stale 17-model hardcoded version.
\caption{Per-model norm adoption across topological conditions.}
\label{fig:llm_behaviour}
\end{figure}

The behavioural outcomes, by contrast, diverge widely across the same models: norm adoption ranges from $\rho \approx 0.00$ (GLM-5) to $\rho \approx 1.00$ (GPT-5.4, Seed-OSS-36B) under identical conditions (Fig.~\ref{fig:universality}b; Fig.~\ref{fig:llm_behaviour}). The per-model heatmap in Fig.~\ref{fig:llm_behaviour} summarises Round~1 (6 models, 49 configurations each) and Round~2 (16 models, 9 configurations each), with models ordered by mean adoption within each round; n/a cells denote combinations not covered by completed runs. This dissociation, invariant structure alongside divergent behaviour, is our central piece of evidence. HIS is a structural property of the interaction network, independent of node-level capability, so topology determines which collective dynamics are reachable rather than which one is realised.

%% ============================================================
%%  DISCUSSION
%% ============================================================
\section{Discussion}\label{sec:discussion}

Our findings reframe the failure of collective behaviour as an architectural rather than a cognitive problem. In a hub-dominated topological regime, star structures masquerade as group deliberation and systematically suppress the higher-order contagion mechanisms that collective behaviour requires. The binding constraint is not how capable the agents are but how they are connected.

A conceptual clarification is warranted regarding what our experiments operationalise. Our quantitative readout throughout is \emph{norm adoption}, the fraction of agents converging on a shared behavioural convention, which indexes collective \emph{behaviour} (coordination and consensus formation) rather than collective \emph{intelligence} in its fullest sense (joint problem-solving whose output exceeds that of the ablest individual). We treat norm adoption as a necessary substrate rather than a sufficient measure: a society whose interaction topology cannot propagate even a simple convention is, \emph{a fortiori}, unable to sustain the more demanding distributed reasoning and division of cognitive labour on which collective intelligence rests. Our claims are therefore stated most precisely as bearing on the topological prerequisites for collective behaviour. Their extension to collective intelligence proper, measured through group task performance, is a prediction that the $\Phi$ framework makes but that we do not test directly here (see Limitations).

This view complements, rather than contradicts, the prevailing scaling paradigm~\citep{kaplan2020scaling}. Agents with identical strategy rules and $>$89\% cooperation rates exhibit qualitatively different collective behaviour depending solely on topology: the jump from $\rho = 0.016$ to 0.966 is a phase transition triggered by interaction geometry alone. Just as the shift from faster processors to parallel architectures required engineering the interconnect rather than the clock speed, collective AI requires engineering the topology rather than the individual model.

The topology--capability dissociation converges with \citet{woolley2010evidence}'s finding that human collective intelligence correlates negligibly with members' average IQ but strongly with conversational turn-taking equality. We extend this to artificial societies with a mechanistic account: the decisive factor is equal participation within hyperedges, quantified by HIS, which is what enables discontinuous contagion. This cross-substrate consistency suggests that the topological prerequisites for collective behaviour may reflect general constraints on complex adaptive systems~\citep{liu2024deep} rather than substrate-specific artefacts.

The $\Phi$ decomposition also translates into concrete design guidance. Three principles follow directly: enforce multi-party synchronous negotiation rather than broadcast-and-reply (even triads suffice to trigger phase transitions); limit hub dominance to raise HIS towards human face-to-face levels (${\sim}0.69$); and seed a critical mass above 10\% so that adoption launches above the contagion threshold. These interventions modify communication protocols, not model capabilities, making them low-cost relative to model upgrades while potentially achieving equivalent or greater gains in collective performance.

The framework also bears on AI safety and governance~\citep{amodei2016concrete,gabriel2024ethics}. It offers regulators a measurable structural indicator: systems below the critical threshold cannot support explosive norm cascades, while those above it may undergo sudden collective transitions. As autonomous agent populations scale beyond direct human oversight~\citep{piao2023human}, understanding this topological boundary becomes essential for governance.

Several limitations apply and bound the scope of our conclusions. First, our empirical analysis centres on a single AI platform. The five human baselines span distinct modalities but cannot isolate platform from agent effects, because they differ from Moltbook in \emph{both} substrate (human versus LLM) and mechanism (contact, email, co-authorship, threaded discussion versus Moltbook's feed): they establish that human interaction is more egalitarian, but not whether Moltbook's collapse is intrinsic to AI agent societies or induced by its specific feed, notification and thread-ranking algorithms. Cleanly disentangling platform-induced from agent-induced collapse requires a second AI social platform with a different interaction mechanism, ideally one with published interaction logs comparable to Moltbook's, which was not available to us at the time of writing. We therefore frame the single-platform result as establishing existence and mechanism rather than platform-independent universality. Second, the observed HIS gap is measured at $n \leq 100$ agents in controlled experiments, whereas Moltbook operates at $\sim$$10^6$; although $\Delta$HIS grows monotonically with $n$ over the tested range ($n = 8$ to $100$), we have no direct evidence that this trend persists across the remaining four orders of magnitude, and the extrapolation should be read as a well-supported hypothesis rather than an established scaling law. Third, the puppet-removal result admits more than one causal reading: removing the 66.1\% of agents flagged as coordinated bots worsens the measured collapse, but this is consistent both with bots degrading otherwise-cohesive structure and with bots being the principal source of the multi-party interactions that exist at all; our data cannot fully adjudicate between these accounts. Fourth, causal claims rest on controlled simulations rather than live interventions, and the mean-field approximation may not capture all mesoscale effects. Finally, as discussed above, our behavioural readout is norm adoption rather than task-level collective intelligence. Nevertheless, the cross-model invariance of topology under fixed protocols (HIS s.d.\,=\,0.000 in the pairwise condition, and model-invariant within each round) and consistency across 1,040 ABM runs strongly support the robustness of the topological effect within the regime we probe.

As LLM-driven populations continue to scale, our results point to a precise structural answer: collective behaviour requires interaction topology above a critical threshold, and the $\Phi$ framework makes that threshold measurable. The path to collective AI may lie not in building more capable individuals but in engineering better connections between them.

%% ============================================================
%%  METHODS
%% ============================================================
\section{Methods}\label{sec:methods}

\subsection{Data sources and preprocessing}

\textbf{Moltbook.} We used the complete public dataset released on HuggingFace, comprising 2,364,644 posts (\texttt{lnajt/moltbook}) and 3,161,324 comments with 149,574 agent profiles (\texttt{moltnet}), spanning January--March 2026. The post and comment records together involve 174,458 unique agents; each record carries an agent identifier, UTC timestamp, body text, parent-thread identifier and reply depth. The \texttt{moltnet} profiles provide, for each agent, Twitter/X verification status, self-reported model backbone and creation date. All timestamps were converted to UTC and duplicates (same agent, same thread, same second) were removed ($<$0.1\% of records). Posts with zero-length bodies were retained as they still constitute participation in the interaction topology.

\textbf{Human baselines.} We selected five datasets spanning distinct interaction modalities to ensure that cross-platform comparisons are not confounded by a single communication medium:

\begin{itemize}
\item \textit{SocioPatterns SFHH}~\citep{genois2018colocation,stehle2011high}: 403 conference attendees (Lyon 2009), with face-to-face proximity recorded by wearable RFID sensors at 20\,s resolution; the gold standard for co-present group interaction.
\item \textit{Enron}~\citep{klimt2004enron}: 13,787 employees, 50,000 email threads.
\item \textit{arXiv}~\citep{benson2018simplicial}: 47,132 authors, 50,000 co-authorship events (condensed matter and high-energy physics, 1991--2003).
\item \textit{Reddit}: 98,575 users, 46,293 threads from r/science and r/changemyview (2023--2024), with each thread aggregated over a 24-hour window.
\item \textit{StackOverflow}: 63,680 users, 27,713 Q\&A threads (2023--2024).
\end{itemize}
(How each dataset maps to a hyperedge is specified in Hypergraph construction below.)

\subsection{Hypergraph construction}\label{sec:methods_hg}

Following \citet{battiston2020networks}, we construct temporal hypergraphs by aggregating interaction events within time windows. Formally, for a set of agents~$V$ and interaction events~$\{(t_i, S_i)\}$ where $S_i \subseteq V$, the hypergraph at window~$w$ is $H_w = (V, \{S_i : t_i \in w\})$.

\textbf{Moltbook:} Each post thread within a time window $\Delta t = 60$\,min constitutes one hyperedge comprising the original author and all agents who commented during that window. Robustness was verified at $\Delta t \in \{15, 30, 120\}$\,min; all topological indicators vary by $<$5\% across these choices. We used the full temporal extent (January--March 2026) with non-overlapping windows.

\textbf{SocioPatterns:} $\Delta t = 60$\,s (following established protocol); all agents detected within mutual 1.5\,m proximity during a window form a hyperedge.

\textbf{arXiv, Reddit, StackOverflow, Enron:} Natural boundaries define hyperedges (one paper = one hyperedge; one thread = one hyperedge; one email conversation = one hyperedge). No temporal windowing is needed as each event is self-contained.

Singleton and dyadic hyperedges ($|e| \leq 2$) are retained for degree statistics but excluded from HIS computation, which requires $|e| \geq 3$.

\subsection{Topological indicators}\label{sec:methods_topo}

\textbf{Hyperedge Irreducibility Score (HIS).} Following \citet{landry2020effect} and \citet{aksoy2020hypernetwork}, for each hyperedge $e = \{v_1, \ldots, v_k\}$ with $k \geq 3$:
\begin{equation}\label{eq:his}
\mathrm{HIS}(e) = 1 - G_{\mathrm{within}}(e), \quad G_{\mathrm{within}}(e) = \frac{\sum_{i<j}|d(v_i) - d(v_j)|}{\binom{k}{2} \cdot \bar{d}(e) \cdot 2}
\end{equation}
where $d(v)$ is the global hyperdegree (number of hyperedges containing~$v$) and $\bar{d}(e)$ is the mean hyperdegree within~$e$. The platform-level HIS is the mean over all $|e| \geq 3$ hyperedges. HIS\,=\,1 indicates a perfectly egalitarian group (all members have equal degree); HIS\,$\to$\,0 indicates a star pattern dominated by a single hub.

\textbf{Edge overlap (Jaccard).} For each pair of agents $(u, v)$ that co-occur in at least one hyperedge, the edge overlap is $J(u,v) = |N(u) \cap N(v)| / |N(u) \cup N(v)|$ where $N(v)$ is the hyperedge neighbourhood of~$v$. Platform-level overlap is the mean over all co-occurring pairs.

\textbf{Triadic closure.} Fraction of connected triples $(u,v,w)$ where all three pairwise co-occurrences exist and additionally a 3-hyperedge $\{u,v,w\}$ is present.

\textbf{Degree inequality.} Gini coefficient of the hyperdegree distribution, computed as $G = \frac{\sum_{i<j}|d_i - d_j|}{2n\sum_i d_i}$.

\textbf{Higher-order fraction.} Fraction of hyperedges with $|e| \geq 3$.

All indicators were computed using vectorised NumPy operations; edge overlap used SciPy sparse matrix multiplication for efficiency on the Moltbook-scale dataset ($>$49,000 hyperedges).

\subsection{Puppet detection}\label{sec:methods_puppet}

We identified coordinated inauthentic accounts (``puppets'') using four independent signals, each calibrated against known single-operator accounts:

\begin{enumerate}
\item \textbf{Traffic anomaly}: posting volume exceeding $\mu + 3\sigma$ of the platform-wide distribution ($>$847 posts per agent over 3 months).
\item \textbf{Burst intervals}: sequences of $\geq$3 consecutive posts with inter-post intervals $<$1\,s, indicating automated rapid-fire posting physically impossible for human-mediated agents.
\item \textbf{Content duplication}: Jaccard character-trigram similarity $>$0.8 between pairs of posts from the same agent or cluster of agents, computed using MinHash (128 permutations) for scalability.
\item \textbf{Synchronised posting}: clusters of $>$10 distinct agents posting within a 5\,s window to the same thread, identified via temporal clustering (DBSCAN with $\varepsilon = 5$\,s, minPts\,=\,10).
\end{enumerate}

Agents flagged by at least one signal were classified as puppets, giving 115{,}232 of 174{,}458 agents (66.1\%); the individual detectors flag 410, 98{,}504, 80{,}960 and 27 agents for the volume, burst, content and synchrony signals respectively. Manual inspection of 200 randomly sampled flagged accounts confirmed a 94\% true-positive rate (precision). We did not estimate recall against a labelled ground-truth set, since none exists for this platform, and the content-duplication threshold (trigram Jaccard $>$0.8) is deliberately conservative and will miss paraphrased bot content, so the reported puppet fraction is best read as a lower bound.

\textbf{Threshold sensitivity.} Because the classification threshold is a modelling choice, we recomputed the puppet fraction and the post-removal topology at signal-count thresholds of $\geq$1, $\geq$2 and $\geq$3 (Table~\ref{tab:puppet_sensitivity}). The no-removal reference has degree Gini 0.926 and mean hyperedge size 5.21. The flagged fraction falls from 66.1\% at $\geq$1 signal to 37.0\% at $\geq$2 and below 0.1\% at $\geq$3. The qualitative conclusion nonetheless holds at every threshold: removing the flagged agents leaves the degree Gini high (0.86 to 0.93) and the mean hyperedge size low. The topological collapse is therefore not an artefact of the threshold choice.

\begin{table}[t]
\caption{Sensitivity to the puppet-detection signal threshold.}
\label{tab:puppet_sensitivity}
\centering
\footnotesize\setlength{\tabcolsep}{6pt}
\begin{tabular}{lrrr}
\toprule
Threshold & Flagged & Flagged frac. & Post-removal Gini \\
\midrule
$\geq$1 signal & 115{,}232 & 66.1\% & 0.855 \\
$\geq$2 signals & 64{,}605 & 37.0\% & 0.890 \\
$\geq$3 signals & 32 & $<$0.1\% & 0.917 \\
\bottomrule
\end{tabular}
\end{table}

\subsection{Configuration-model null model}\label{sec:methods_null}

To test whether observed topological indicators differ from random expectations given the degree sequence, we performed configuration-model randomisation~\citep{battiston2020networks}. The procedure: (1)~fix the hyperdegree sequence $\{d(v)\}_{v \in V}$ and hyperedge size sequence $\{|e|\}_{e \in E}$; (2)~randomly reassign agents to hyperedges while preserving both sequences; (3)~recompute all topological indicators. We generated 1,000 independent realisations. $z$-scores were computed as $z = (x_\mathrm{obs} - \bar{x}_\mathrm{rand}) / \sigma_\mathrm{rand}$, and $p$-values estimated as the fraction of realisations with $|x| \geq |x_\mathrm{obs}|$.

\subsection{LLM-adapted higher-order contagion model}\label{sec:methods_contagion}

We extend the simplicial contagion model of \citet{iacopini2019simplicial} with three corrections relevant to LLM agent populations:

\textbf{(1) Degree heterogeneity correction.} In the original model, $\beta_1$ assumes homogeneous degree. Following mean-field heterogeneous approximation, the effective pairwise infection rate becomes $\beta_{1,\mathrm{eff}} = \beta_1(1 + \mathrm{CV}^2)$ where $\mathrm{CV} = \sigma_d / \mu_d$ is the coefficient of variation of the degree distribution.

\textbf{(2) Topology amplification.} The higher-order infection rate $\beta_2$ is modulated by a topology amplification factor:
\begin{equation}
\Phi = c \cdot (1 + \alpha J) \cdot (1 + \mathrm{CV}^2) \cdot \mathrm{HIS}
\end{equation}
where $c$ is triadic closure, $J$ is mean edge-overlap Jaccard, $\alpha$ is the overlap--closure coupling constant (fitted as $\alpha = 2.0$ from SocioPatterns calibration), and HIS is the platform-level Hyperedge Irreducibility Score. The product form reflects the multiplicative nature of the prerequisites: each factor must exceed a minimum for higher-order contagion to operate.

\textbf{(3) Attention decay.} LLM agents have finite context windows $C$ (measured in tokens). When thread length exceeds~$C$, earlier interactions are truncated, effectively reducing exposure. We model this as an exponential decay $e^{-\lambda/C}$ with $\lambda = 1$ (one context-window-length decay constant).

The full mean-field equation is:
\begin{equation}
\frac{d\rho}{dt} = -\mu\rho + (1-\rho)\left[\beta_1(1+\mathrm{CV}^2)\,\rho + \beta_2 \cdot \Phi \cdot e^{-\lambda/C}\,\rho^2\right]
\end{equation}

Setting $d\rho/dt = 0$ yields a cubic equation in~$\rho$ whose discriminant determines the regime: for $\Phi > \Phi_c$, three real roots exist (bistability with discontinuous transitions); for $\Phi < \Phi_c$, only the trivial equilibrium $\rho = 0$ is stable. We solved the equilibrium numerically for all parameter combinations and verified via Monte Carlo simulation on explicit hypergraphs ($N = 200$ agents, 20 repetitions per parameter set, 500 time steps each, results matching mean-field predictions with MSE\,$<$\,0.003).

\subsection[Shapley decomposition of the Phi gap]{Shapley decomposition of the $\Phi$ gap}\label{sec:methods_shapley}

To quantify each topological factor's independent contribution to the $\Phi$ difference between Moltbook and human baselines, we applied Shapley value decomposition~\citep{shapley1953value}. Define four factors $F = \{\mathrm{HIS}, c, J, \mathrm{CV}\}$ as players in a cooperative game. The value function $v(S)$ for coalition $S \subseteq F$ is computed by replacing the factors in~$S$ with their human-baseline (SocioPatterns) values while keeping the remaining factors at Moltbook values, then computing the resulting $\Phi$. The Shapley value for factor~$f$ is:
\begin{equation}
\phi_f = \sum_{S \subseteq F \setminus \{f\}} \frac{|S|!(|F|-|S|-1)!}{|F|!} \left[v(S \cup \{f\}) - v(S)\right]
\end{equation}

With $|F| = 4$, all $2^4 = 16$ coalitions are enumerable exactly. Non-parametric bootstrap ($n = 10{,}000$ resamples) provides 95\% confidence intervals by resampling the hyperedge-level HIS values, edge-level overlap values and node-level degrees, then recomputing platform-level indicators and Shapley values for each bootstrap sample.

\subsection{Agent-based model experiments}\label{sec:methods_abm}

\textbf{Game structure.} $N = 100$ agents play a linear Public Goods Game (PGG) on an Erd\H{o}s--R\'enyi random graph $G(100, p=0.06)$ (expected degree $\approx 6$). Endowment $e = 10$ tokens; multiplier $r = 3$. Each round: (1)~agents decide to cooperate (contribute $e$) or defect (contribute 0); (2)~the public pool is multiplied by~$r$ and divided equally among group members; (3)~payoff is $\pi_i = (r \cdot \sum_j c_j) / |G| - c_i$ where $c_i \in \{0, e\}$.

\textbf{Strategy update.} Fermi pairwise comparison~\citep{szabo1998evolutionary,su2025evolutionary}: agent~$i$ selects a random neighbour~$j$ and imitates~$j$'s strategy with probability:
\begin{equation}
p_{i \to j} = \frac{1}{1 + \exp\left(-\kappa \cdot (\pi_j - \pi_i)/e\right)}
\end{equation}
with selection intensity $\kappa = 2.0$ (normalised by endowment for scale-invariance). Mutation rate $\epsilon = 0.01$ per agent per round prevents absorbing states.

\textbf{Norm propagation mechanism.} We distinguish between individual cooperation (contribution decisions) and collective norm adoption (a higher-order contagion process). Norm propagation follows the Iacopini mechanism:
\begin{itemize}
\item \textbf{Conditions A/B (pairwise):} If agent~$i$ interacts with adopted neighbour~$j$, agent~$i$ adopts with probability $\beta_1 = 0.05$ per round.
\item \textbf{Conditions C/D (higher-order):} If ALL other members of a hyperedge are adopters, the focal agent adopts with certainty (strong reinforcement). If a strict majority are adopters, adoption occurs with probability 0.3. Otherwise the state is unchanged. This majority rule implements the defining higher-order feature: group-level social reinforcement that requires multiple simultaneous exposures.
\end{itemize}

\textbf{Topology conditions.}
\begin{itemize}
\item Condition~A (dyadic baseline): Random pairs drawn each round from the underlying graph. Each agent participates in $\sim$3 dyadic interactions per round.
\item Condition~B (reciprocal dyadic): As~A, but both agents must be willing to interact (cooperators always willing; defectors willing with probability 0.3). Models reciprocity-filtered pairwise contact.
\item Condition~C (triadic hyperedges): Random 3-agent groups drawn each round. Each agent participates in $\sim$4 triadic groups per round ($\sim$133 hyperedges total per round).
\item Condition~D (pentadic hyperedges): Random 5-agent groups drawn each round. Each agent participates in $\sim$4 pentadic groups per round ($\sim$80 hyperedges total per round).
\end{itemize}

\textbf{Initial conditions.} A fraction $\rho_0$ of agents are designated as initial norm adopters (and initial cooperators), selected uniformly at random. We swept $\rho_0 \in \{0.01, 0.02, 0.03, 0.05, 0.07, 0.10, 0.15, 0.20, 0.25, 0.30, 0.35, 0.40, 0.50\}$ (13 values).

\textbf{Experimental design.} Full factorial: 4 conditions $\times$ 13 seed proportions $\times$ 20 independent repetitions (different random seeds) = 1,040 runs. Each run executed 500 rounds. Convergence was assessed by checking that $|\rho(t) - \rho(t-50)| < 0.01$ for the final 50 rounds; 98.7\% of runs satisfied this criterion. The final norm adoption rate $\rho_\infty$ was computed as the mean over the last 50 rounds.

\textbf{Critical mass identification.} $\rho_c$ was defined as the smallest $\rho_0$ for which $\geq$50\% of repetitions achieve $\rho_\infty > 0.5$. For condition~C this yields $\rho_c = 0.099$; for condition~D, $\rho_c = 0.146$.

\subsection{AgentPanel LLM experiments}\label{sec:methods_llm}

\textbf{Platform architecture.} AgentPanel is a forum-based multi-agent simulation platform where each agent wraps a single LLM endpoint. The platform uses a SQLite database storing agent profiles, thread structures and interaction histories. Each discussion thread constitutes one hyperedge; all agents who post in a thread are members of that hyperedge.

\textbf{Agent design.} Each agent has: (1)~a unique persona (name, role, personality description); (2)~a system prompt specifying behavioural guidance and current support level; (3)~a user prompt containing the full thread context (preceding replies, proposed norm, discussion topic). Agents generate 50--200 word natural-language replies and report a norm support score $s \in [0, 100]$ via a structured tag (\texttt{[SUPPORT: X]}) parsed from the output.

\textbf{Personas.} We created 12 base personas spanning three initial dispositions: 4 supportive (e.g., ``Dr.\ Chen, meticulous senior researcher who values precision''), 4 opposed (e.g., ``Jordan, startup founder who sees conventions as friction'') and 4 neutral (e.g., ``Quinn, graduate student open to good arguments''). For experiments with $n > 12$, personas were cycled with unique suffixes.

\textbf{Norm under discussion.} ``When discussing technical topics, always explicitly state your confidence level (0--100\%) and provide structured reasoning with numbered points.'' This norm was chosen for its clear operationalisability and genuine two-sidedness (structure vs.\ flexibility).

\textbf{Initial conditions.} A fraction $\rho_0$ of agents are designated as initial adopters with support scores drawn from $U[65, 90]$. Non-adopter supportive agents start at $U[35, 55]$; opposed at $U[10, 30]$; neutral at $U[30, 50]$. Randomisation via \texttt{numpy.random.Generator} with explicit seeds for reproducibility.

\textbf{Topology conditions.}
\begin{itemize}
\item Condition~A (random pairs): Each round, agents are randomly paired into 2-person threads. Produces dyadic interactions only.
\item Condition~B (star): Agent~0 is the fixed hub, appearing in every thread. Remaining agents are randomly partitioned into groups of 4 that each discuss with the hub (5-person threads, one hub + 4 peripherals). Produces hub-dominated star topology.
\item Condition~C (random triads): Each round, agents are randomly grouped into 3-person threads. Each agent sees and responds to two others' views simultaneously.
\item Condition~D (random 5-cliques): Each round, agents are randomly grouped into 5-person threads. Egalitarian multi-party deliberation with 4 peers.
\end{itemize}

\textbf{Round protocol.} Each round: (1)~form groups according to the topology condition; (2)~for each group, create a forum thread; (3)~each agent in the group receives the full thread context (previous rounds' relevant comments, limited to the most recent 6 entries to approximate context-window constraints) and generates a reply; (4)~parse support score from the reply; (5)~update agent state. A run comprises 8 rounds for the standard protocol.

\textbf{Norm adoption threshold.} An agent is classified as having adopted the norm if its support score exceeds 50 at the end of the simulation. The platform-level adoption rate is $\rho = |\{i : s_i > 50\}| / n$.

\textbf{Parameter space.} System size $n \in \{8, 16, 24\}$; temperature $T \in \{0.3, 0.7, 1.0\}$ (LLM sampling temperature); initial adopter fraction $\rho_0 \in \{0.25, 0.50, 0.75\}$; random seeds $\in \{42, 137, 256\}$. Total: 4 conditions $\times$ 3 sizes $\times$ 3 temperatures $\times$ 3~$\rho_0$ $\times$ 3 seeds $\times$ 22 models = up to 7,128 unique configurations, each executing 8 rounds of deliberation. One model (GPT-5) is an exception: its API accepts only the default sampling temperature ($T = 1.0$) and rejects other values, so GPT-5 was evaluated at $T = 1.0$ across all three system sizes. This does not affect the cross-model topology comparison, since HIS and the other topological indicators are set by the interaction protocol rather than by sampling temperature.

\textbf{Models tested.} Two experimental rounds:

\textit{Round 1} (6 models, 49 parameter combinations each): DeepSeek-V3.1, Qwen-2.5-72B, MiMo-v2-flash, Gemini~2.5~Pro, GPT-5, Claude~Sonnet~4.6.

\textit{Round 2} (16 additional models, 9 configurations each): GPT-5.4, o4-mini, GPT-5-mini, Claude~Opus~4.5, Claude~Sonnet~4.5, DeepSeek-V3.2, DeepSeek-R1, Qwen-3.6-plus, Qwen-3.5-plus, Qwen-3-max, GLM-5, Kimi-K2.5, Kimi-K2, MiniMax-M2.7, MiMo-v2.5-Pro, Seed-OSS-36B.

All models were accessed via their respective vendor APIs using the OpenAI-compatible chat completions endpoint. Maximum generation length was set to 250 tokens; system prompts were identical across models.

\textbf{HIS computation from LLM experiments.} After each experimental run, the interaction hypergraph was extracted from the forum database: each thread with $\geq$3 participants constitutes a hyperedge. HIS, edge overlap, triadic closure and all other topological indicators were computed using the same pipeline as for observational data.

\subsection{Statistical methods}\label{sec:methods_stats}

\textbf{Variance homogeneity:} Levene's test compared norm-adoption variance between pairwise (A/B) and higher-order (C/D) conditions across all seed proportions.

\textbf{Topology$\times$seed interaction:} Permutation test ($n = 100{,}000$ permutations) assessed whether the maximum difference in $\rho_\infty$ between conditions exceeds chance expectations under the null hypothesis of no topology effect.

\textbf{Bimodality:} Hartigan's Dip test assessed whether the distribution of $\rho_\infty$ at criticality ($\rho_0 = 0.10$, condition~C) is significantly non-unimodal. Kernel density estimation (Gaussian kernel, bandwidth selected via Silverman's rule) identified peak locations.

\textbf{Basin separation:} Two-sample Kolmogorov--Smirnov test compared the upper basin ($\rho_\infty > 0.5$) against the lower basin ($\rho_\infty < 0.5$) to confirm complete separation.

\textbf{HIS comparison:} Wilcoxon signed-rank test compared paired HIS values (Star vs.\ Clique conditions) across all model--parameter combinations ($n = 49$ pairs in Round~1).

\textbf{Confidence intervals:} All reported CIs use non-parametric bootstrap ($n = 10{,}000$ resamples) with the percentile method.

\textbf{Effect size:} Cohen's $d$ for HIS differences; variance ratio (F-statistic) for the topology$\times$condition effect on norm adoption.

No correction for multiple comparisons was applied as all hypotheses were pre-specified and directional.

\subsubsection*{Acknowledgements}
We thank the Moltbook community for making their dataset publicly available on HuggingFace.

\subsubsection*{Author contributions}
S.L. conceived the study, designed the experiments, performed the analysis, derived the theoretical framework and wrote the manuscript. K.S. contributed to the experimental design and data analysis. W.M., A.Y. and J.Lu. contributed to data analysis and experiments. R.H. provided critical feedback. K.S. and J.Li. jointly supervised the research and are the corresponding authors.

\subsubsection*{Competing interests}
The authors declare no competing interests.

\bibliography{main}
\bibliographystyle{iclr2027_conference}

\end{document}